\documentclass[floats, eqnum, showpacs, nofootinbib, 
preprint,
eqsecnum]{revtex4-1}
\usepackage{color,graphicx}
\usepackage{amsfonts}
\usepackage{amssymb}
\begin{document}

\title{Strong deflection limit analysis and gravitational lensing of an Ellis wormhole}

\author{Naoki Tsukamoto}\email{tsukamoto@rikkyo.ac.jp}

\affiliation{
School of Physics, Huazhong University of Science and Technology, Wuhan 430074, China
}
\date{\today}

\begin{abstract}
Observations of gravitational lenses in strong gravitational fields give us a clue to understanding dark compact objects.
In this paper, we extend a method to obtain a deflection angle in a strong deflection limit provided by Bozza~[Phys. Rev. D 66, 103001 (2002)]
to apply to ultrastatic spacetimes.
We also discuss on the order of an error term in the deflection angle.
Using the improved method, we consider gravitational lensing by an Ellis wormhole, which is an ultrastatic wormhole of the Morris-Thorne class.
\end{abstract}


\preprint{
}

\maketitle

\section{Introduction}
Gravitational lensing is a useful tool to find massive and dark objects such as extrasolar planets. 
For the last one hundred years, gravitational lenses under weak-field approximations have been investigated eagerly.
See~\cite{Schneider_Ehlers_Falco_1992,Petters_Levine_Wambsganss_2001,Perlick_2004_Living_Rev,Schneider_Kochanek_Wambsganss_2006}
for the details of gravitational lensing.
Gravitational lenses in strong gravitational fields also have been studied because
observations of gravitational lenses in strong gravitational fields give us information of compact objects like black holes.
Gravitational lensing in a strong gravitational field was pioneered by Darwin~\cite{Darwin_1959}.
Gravitational lensing effects of light rays that are emitted by a source and wind around a compact lens object arbitrary times on a light sphere in a strong gravitational field
have been revived several times~\cite{Atkinson_1965,Luminet_1979,Ohanian_1987,Nemiroff_1993,Virbhadra_Ellis_2000,Bozza_2002,Hasse_Perlick_2002,Bozza:2007gt,Bozza:2008ev,Bozza_2010}.

Wormholes also cause gravitational lensing effects in both weak and strong gravitational fields.
Gravitational lenses by wormholes were investigated first 
by~Kim and Cho~\cite{Kim_Cho_1994} and Cramer \textit{et al.}~\cite{Cramer_Forward_Morris_Visser_Benford_Landis_1995}
and then gravitational lensing effects of negative mass wormholes~\cite{Takahashi_Asada_2013,Safonova_Torres_Romero_2001_Jan} 
and positive mass wormholes~\cite{Rahaman_Kalam_Chakraborty_2007,Tejeiro_Larranaga_2012,Nandi_Zhang_Zakharov_2006,Dey_Sen_2008} were studied.

A massless wormhole called Ellis wormhole was found by Ellis as the simplest wormhole solution of the Einstein equations with a phantom scalar field in 1973~\cite{Ellis_1973}
and by Bronnikov in a scalar-tensor theory in the same year~\cite{Bronnikov_1973}.
The Ellis wormhole metric is often referred to as the simplest Morris-Thorne wormhole~\cite{Morris_Thorne_1988,Morris_Thorne_Yurtsever_1988}.
Instability of Ellis wormhole spacetimes was reported in~\cite{Shinkai_Hayward_2002,Nandi:2016ccg}, contrary to a conclusion of Ref.~\cite{Armendariz-Picon_2002}.
Several wormhole solutions with the Ellis wormhole metric as their simplest cases 
and with some exotic matters as a source of its metric but not the phantom scalar field
are known~\cite{Kar:2002xa,Das:2005un,Shatskiy:2008us,Myrzakulov:2015kda}.
The stability of the wormholes depends on the matters.
Recently, in Ref.~\cite{Bronnikov:2013coa}, Bronnikov \textit{et al.} found a linearly stable wormhole 
with the Ellis wormhole metric that is filled with electrically charged dust with negative energy density~\cite{Shatskiy:2008us,Novikov:2012uj}
under both spherically symmetric and axial perturbations.
The quasinormal modes have been investigated by Konoplya and Zhidenko~\cite{Konoplya:2016hmd}.

A deflection angle of a light ray in the Ellis wormhole spacetime was studied first by Chetouani and Clement~\cite{Chetouani_Clement_1984} 
and later in Ref.~\cite{Nandi_Zhang_Zakharov_2006,Dey_Sen_2008,Muller:2008zza,Bhattacharya:2010zzb,Gibbons_Vyska_2012,Nakajima_Asada_2012,Tsukamoto_Harada_Yajima_2012}.
Micro lens~\cite{Abe_2010,Kitamura_Nakajima_Asada_2013,Tsukamoto:2014dta,Lukmanova_2016}, 
astrometric image centroid displacements~\cite{Toki_Kitamura_Asada_Abe_2011,Kitamura_Izumi_Nakajima_Hagiwara_Asada_2013},
the Einstein rings~\cite{Tsukamoto_Harada_Yajima_2012},
the time delay of light rays~\cite{Nakajima:2014nba},
the signed magnification sum~\cite{Tsukamoto_Harada_2013},
the gravitational lensing shear~\cite{Izumi_Hagiwara_Nakajima_Kitamura_Asada_2013},
constraints of the number density from gravitational lensing observations~\cite{Takahashi_Asada_2013,Yoo_Harada_Tsukamoto_2013},
a wave effect of gravitational lenses~\cite{Yoo_Harada_Tsukamoto_2013},
shadows surrounded by a plasma~\cite{Perlick:2015vta} and optically thin dust~\cite{Ohgami:2015nra},
binary gravitational lenses~\cite{Bozza:2015wbw},
a particle collision~\cite{Tsukamoto:2014swa},
and several observables such as rotation curves~\cite{Bozza:2015haa}
in the Ellis wormhole spacetime 
and in general spacetimes that are coincident with the Ellis wormhole spacetime under the weak-field approximation
have been investigated.
The visualization of the Ellis wormhole was studied by Muller~\cite{Muller_2004}.
The gravitational lenses of a source on the other side of the Ellis wormhole 
were investigated by Perlick~\cite{Perlick_2004_Phys_Rev_D} and Tsukamoto and Harada~\cite{Tsukamoto:2016zdu}.

In this paper, we consider a deflection angle in a strong deflection limit in an Ellis wormhole spacetime.
We use a behavior of the complete elliptic integral of the first kind to obtain it 
and then we cross-check it by using a well-known method 
to obtain the deflection angle in the strong deflection limit in a general spherical symmetric spacetime investigated by Bozza~\cite{Bozza_2002}.
For cross-checking, we give two small revisions of the method.
First, we correct the order of an error term of the deflection angle.
Second, we point out that the method does not work in ultrastatic spacetimes such as the Ellis wormhole spacetime
and we extend the method to apply to the Ellis wormhole spacetime.

A total magnification theorem that the total magnification of 
lensed images by an isolated mass is always larger than unity under the weak-field approximation~\cite{Petters_Levine_Wambsganss_2001} is known well. 
Recently, Abe found that the microlensing light curves in an Ellis wormhole spacetime can be demagnified near the peaks 
under the weak field approximation~\cite{Abe_2010}.
This result apparently violates the total magnification theorem 
but it does not contradict the theorem since the Ellis wormhole has zero Arnowitt-Deser-Misner~(ADM) masses.

In~\cite{Abe_2010}, Abe considered that light rays passing by a light sphere does not contribute to the microlensing light curves because of the absence of
the light sphere of the Ellis wormhole. 
An Ellis wormhole, however, has a light sphere and light rays passing by the light sphere may contribute to the microlensing light curves 
and they may disturb the characteristic gutters of the light curves near the peak.
In the Schwarzschild spacetime, images of light rays passing by the light sphere are much fainter than the images under the weak-field approximation 
and the former images do not affect the microlensing light curves~\cite{Bozza:2001xd,Petters:2002fa}.
How about the Ellis wormhole spacetime case?
Since an Ellis wormhole has vanishing ADM masses, the asymptotic behavior of light rays is qualitatively different form ones in the Schwarzschild spacetime.
Tsukamoto \textit{et al.}~\cite{Tsukamoto_Harada_Yajima_2012} pointed out 
that the radius of the Einstein ring made by light rays in the weak-field approximation in an Ellis wormhole spacetime 
is very different from the one in the Schwarzschild spacetime 
if we assume that the sizes of their light spheres are same.
Thus, the ratio of the total magnifications of images under the weak-field approximation and in the strong deflection limit in the Ellis wormhole spacetime
is qualitatively different from the one in the the Schwarzschild spacetime.
In this paper, we discuss the effect of images in the strong deflection limit on microlensing light curves in the Ellis wormhole spacetime 
as an application of our result on the deflection angle in the strong deflection limit.
 
This paper is organized as follows. 
In Sec.~II we review the deflection angle in the Ellis wormhole spacetime 
and obtain the one in a strong deflection limit.
In Sec.~III, we extend a strong deflection limit analysis and examine the deflection angle in the strong deflection limit.
In Sec.~IV we introduce a lens equation.
In Sec.~V we calculate observables in the strong deflection limit.
In Sec.~VI, we show the effect of images in the strong gravitational field on light curves.
We discuss our results in Sec.~VII.
In Appendix~A, we show that the Ellis wormhole has vanishing ADM masses.
In Appendix~B, we compare the deflection angle obtained in the current paper with ones in previous works.
In this paper we use the units in which the light speed $c$ and Newton's constant $G$ are unity.

\section{Deflection angle in an Ellis wormhole spacetime in a strong deflection limit}
In this section, we obtain a deflection angle $\alpha$ of light rays in an Ellis wormhole spacetime in a strong deflection limit given by the following form,
\begin{equation}\label{eq:Strong}
\alpha(b)
=-\bar{a} \log \left( \frac{b}{b_{c}}-1 \right) + \bar{b} +O\left( \left( b-b_{c} \right) \log\left(b-b_{c}\right) \right),
\end{equation}
where $\bar{a}$ is a positive constant, $\bar{b}$ is a constant, $b$ is the impact parameter, and $b_{c}$ is the critical impact parameter~\cite{Bozza_2002}.  

The line element in the Ellis wormhole spacetime~\cite{Ellis_1973,Bronnikov_1973} is described by
\begin{equation}\label{eq:line_element}
ds^{2}=-dt^{2}+dr^{2}+(r^{2}+a^{2})(d\theta^{2}+\sin^{2}\theta d\phi^{2}),
\end{equation}
where the coordinates are defined in the range $-\infty < t < \infty$, $-\infty < r < \infty$, $0 \leq \theta \leq \pi$, and  $0 \leq \phi < 2\pi$  
and $a$ is a positive constant.
The time translational and axial Killing vectors $t^{\alpha}\partial_{\alpha}=t^{t}$ and $\phi^{\alpha}\partial_{\alpha}=\phi^{\phi}$ exist 
because of stationarity and axisymmery, respectively.
Note that the Ellis wormhole spacetime is not only static but also ultrastatic, i.e., $g_{tt}=\mathrm{constant}$. 
The wormhole throat is at $r=0$.
In the limit $a \rightarrow 0$, the line element is apparently coincident with the one in the Minkowski spacetime.
The Ellis wormhole has vanishing ADM masses in the both sides of the throat
while light rays can deflect because of $a$.
(See Appendix~A for the calculation of its ADM masses.) 
We assume $\theta=\pi/2$ without loss of generality because of spherical symmetry.
We concentrate on a region $r\geq 0$ 
since we are interested in gravitational lenses of light rays which are emitted at a source, 
are scattered by an Ellis wormhole, and reach an observer on the same side of the wormhole. 

The trajectory of photons is described by 
\begin{equation}\label{eq:k_null}
k^{\mu}k_{\mu}=0, 
\end{equation}
where $k^{\mu}$ is the photon wave number.
Equation~(\ref{eq:k_null}) can be expressed by
\begin{equation}\label{eq:TrajectoryPhoton}
\frac{1}{(r^{2}+a^{2})^{2}} \left( \frac{dr}{d\phi} \right)^{2}=\frac{1}{b^{2}}-\frac{1}{r^{2}+a^{2}},
\end{equation}
where $b\equiv L/E$ is the impact parameter and 
$E\equiv -g_{\mu\nu}t^{\mu}k^{\nu}$ and $L\equiv g_{\mu\nu}\phi^{\mu}k^{\nu}$ 
are the conserved energy and angular momentum of the photon, respectively.
We assume $L\geq 0$ and then $b\geq 0$.
The photon is scattered if $b > a$ 
while it falls into the throat if  $b < a$.
In the limit $b \rightarrow b_{c} \equiv a$, the light ray winds around the throat at $r=0$.
The throat is coincident with a light sphere.
We only consider the scatter case, i.e., $b > a$.

From $dr/d\phi=0$ and Eq.~(\ref{eq:TrajectoryPhoton}),
the closest distant is given by $r_{0}=\sqrt{b^{2}-a^{2}}$.
From Eq.(\ref{eq:TrajectoryPhoton}), we obtain the deflection angle of lights as
\begin{equation}
\alpha=2\int^{\infty}_{r_{0}}\frac{bdr}{\sqrt{(r^{2}+a^{2})(r^{2}+a^{2}-b^{2})}}-\pi.
\end{equation}
Introducing $x\equiv b/\sqrt{r^{2}+a^{2}}$ and $k\equiv a/b$,
the deflection angle can be rewritten into
\begin{equation}\label{eq:deflection}
\alpha=2K \left( k \right)-\pi,
\end{equation}
where $K(k)$ is the complete elliptic integral of the first kind which is defined as
\begin{equation}
K(k)=\int^{1}_{0}\frac{dx}{\sqrt{(1-x^{2})(1-k^{2}x^{2})}},
\end{equation}
where $0 < k < 1$.  
The deflection angle~(\ref{eq:deflection}) can be expressed as
\begin{equation}\label{eq:expanded_deflection}
\alpha=\pi \sum^{\infty}_{n=1} \left[ \frac{(2n-1)!!}{(2n)!!} \right]^{2} k^{2n},
\end{equation}
where $!!$ denotes the double factorial.
Under the weak-field approximation $a\ll b$, the deflection angle becomes
\begin{equation}\label{eq:weak_deflection}
\alpha=\frac{\pi}{4} k^{2} +\frac{9\pi}{64} k^{4} +O\left( k^{6} \right).
\end{equation}
and we can transform it into 
\begin{equation}\label{eq:weak_deflection2}
\alpha=\frac{\pi a^{2}}{4r_{0}^{2}}-\frac{7\pi a^{4}}{64r_{0}^{4}}+O\left( \frac{a^{6}}{r^{6}_{0}} \right).
\end{equation}
In Appendix~B, we compare Eq.~(\ref{eq:deflection}) and deflection angles in previous works.

From Eq.~(10) in section~13.~8 in Ref.~\cite{Erdelyi}, in the strong deflection limit $k\rightarrow 1$, 
$K(k)$ becomes 
\begin{equation}
\lim_{k\rightarrow 1}K(k)=-\frac{1}{2} \log(1-k) +\frac{3}{2}\log 2  +O((1-k)\log(1-k)).
\end{equation}
Thus, the deflection angle in the strong deflection limit $b\rightarrow b_{c}=a$ is given by
\begin{equation}\label{eq:strong_field_limit}
\alpha(b)
=-\log \left( \frac{b}{b_{c}}-1 \right) + 3\log 2 -\pi +O((b-b_{c})\log(b-b_{c})).
\end{equation}
Therefore, we get $\bar{a}=1$ and $\bar{b}=3\log 2-\pi$.

\section{Revision of a strong deflection limit analysis and examination of the deflection angle}
We recalculate the deflection angle in the strong deflection limit~(\ref{eq:strong_field_limit}).
We use a method investigated by Bozza~\cite{Bozza_2002} 
to obtain the deflection angle in the strong deflection limit 
in a general static spherically symmetric spacetime 
with a line element
\begin{equation} 
ds^{2}=-A(r)dt^{2}+B(r)dr^{2}+C(r)(d\theta^{2}+\sin\theta^{2}d\phi^{2}),
\end{equation}
where 
\begin{eqnarray}\label{eq:A}
\lim_{r\rightarrow \infty} A(r)&=&1-\frac{2M}{r}+O\left( \frac{1}{r^{2}} \right),\\ \label{eq:B}
\lim_{r\rightarrow \infty} B(r)&=&1+\frac{2M}{r}+O\left( \frac{1}{r^{2}} \right) ,\\ \label{eq:C}
\lim_{r\rightarrow \infty} C(r)&=&r^{2} +O(r),
\end{eqnarray}
where $M$ is the ADM mass.
The Ellis wormhole spacetime has $A(r)=B(r)=1$ and $C(r)=r^{2}+a^{2}$.
The assumptions~(\ref{eq:A})-(\ref{eq:C}) seem to be sufficient but not necessary 
to obtain the deflection angle in the strong deflection limit.
\footnote{Tsukamoto \textit{et al.}~\cite{Tsukamoto:2014dta} 
showed that Bozza's method works also in the Tangherlini spacetime 
with
\begin{eqnarray}
\lim_{r\rightarrow \infty} A(r)&=&1-\frac{m}{r^{n}}+O\left( \frac{1}{r^{n+1}} \right),\\ 
\lim_{r\rightarrow \infty} B(r)&=&1+\frac{m}{r^{n}}+O\left( \frac{1}{r^{n+1}} \right),\\ 
\lim_{r\rightarrow \infty} C(r)&=&r^{2} +O(r),
\end{eqnarray}
where $m$ is a positive constant which is proportional to the mass and $n$ is an arbitrary positive integer.}

We realize that the method does not work right 
because we cannot define a variable $z'$ given by 
\begin{eqnarray}\label{eq:z'}
z'\equiv \frac{A(r)-A(r_{0})}{1-A(r_{0})}.
\end{eqnarray}
See Eqs.~(10) and (11) in ~\cite{Bozza_2002}. 
We revise the method and obtain the deflection angle in the strong deflection limit.
\footnote{If we transform the coordinate $t$ into $c_{t}t$, where $c_{t}$ is  a positive constant but it is not unity,
we can define $z'$ as Eq.~(\ref{eq:z'}). However, $z'$ always vanishes and it is not suitable for our purpose. 
If a spacetime is ultrastatic, we always face the trouble.
Thus, we revise the variable to apply Bozza's method to the Ellis wormhole spacetime.} 
Instead of $z'$, we use a variable $z$ defined as 
\begin{eqnarray}\label{eq:z}
z\equiv 1-\frac{b}{\sqrt{r^{2}+a^{2}}}.
\end{eqnarray}
Using $z$, $2K(k)$ can be expressed by
\begin{equation}
2K(k)=\int^{1}_{0}f(z,k)dz,\end{equation}
where 
\begin{equation}
f(z,k) \equiv \frac{2}{\sqrt{c_{1}(k)z+c_{2}(k)z^{2}-4k^{2}z^{3}+k^{2}z^{4}}},
\end{equation}
where $c_{1}(k) \equiv 2(1-k^{2})$ and $c_{2}(k) \equiv -1+5k^{2}$.
In the strong deflection limit $k\rightarrow 1$, $c_{1}(k)$ and $c_{2}(k)$ become
\begin{eqnarray}\label{eq:c_1}
c_{1}(k) &=& 4(1-k) +O\left((1-k)^{2}\right),\\ \label{eq:c_2}
c_{2}(k) &=& 4+ O(1-k).
\end{eqnarray}
In the strong deflection limit $k\rightarrow 1$, 
the leading order of the divergence in $f(z,k)$ is $z^{-1}$ since $c_{1}(k) \rightarrow 0$.

We divide $2K(k)$ into a divergent part $I_{D}(k)$ and a regular part $I_{R}(k)$, 
\begin{equation}
2K(k)=I_{D}(k)+I_{R}(k).
\end{equation}
The divergent part $I_{D}(k)$ is defined as
\begin{equation}
I_{D}(k) \equiv \int^{1}_{0} f_{0}(z,k)dz,
\end{equation}
where
\begin{equation}
f_{0}(z,k) \equiv \frac{2}{\sqrt{c_{1}(k) z +c_{2}(k) z^{2}}}.
\end{equation}
We can integrate the divergent part $I_{D}(k)$ and obtain 
\begin{equation}
I_{D}(k)
=\frac{4}{\sqrt{-1+5k^{2}}}\log\frac{\sqrt{-1+5k^{2}}+\sqrt{1+3k^{2}}}{\sqrt{2(1-k)(1+k)}}.
\end{equation}
The divergent part $I_{D}(k)$ in the strong deflection limit is obtained by
\begin{equation}
I_{D}(k)=-\log(1-k) +2\log 2 +O((1-k)\log(1-k)).
\end{equation}
The regular part $I_{R}(k)$ is defined by
\begin{equation}
I_{R}(k) \equiv \int^{1}_{0}g(z,k)dz,
\end{equation}
where 
\begin{equation}
g(z,k) \equiv f(z,k)- f_{0}(z,k).
\end{equation}
Since 
\begin{equation}
\lim_{k\rightarrow 1} g(z,k) =\frac{1}{2-z} +O((1-k)\log(1-k)),
\end{equation}
$I_{R}(k)$ in the strong deflection limit $k\rightarrow 1$ is obtained as
\begin{equation}
I_{R}(k)=\log 2 +O((1-k)\log(1-k)).
\end{equation}
Thus, the deflection angle in the strong deflection limit is given by
\begin{equation}\label{eq:strong_field_limit2}
\alpha(k)=-\log(1-k) +3\log 2 -\pi +O((1-k)\log(1-k)). 
\end{equation}
This is equal to Eq.~(\ref{eq:strong_field_limit}).

\section{Lens equation}
We consider that a light ray emitted by a source bends with the deflection angle $\alpha$ near an Ellis wormhole as a lens object
and reaches an observer. 
See Fig.~1.
The observer sights an image with an image angle $\theta$ but not a source with a source angle $\phi$.
We define an effective deflection angle as
\begin{equation}\label{eq:deflection_angle_separeted}
\bar{\alpha}\equiv \alpha-2\pi N,
\end{equation}
where $N$ is a non-negative integer that denotes the winding number of the light ray.
\begin{figure}[htbp]
\begin{center}
\includegraphics[width=80mm]{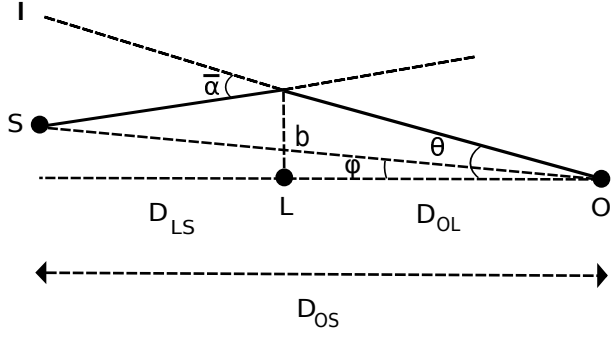}
\end{center}
\caption{Configuration of a gravitational lens.
A light ray emitted by a source $S$ bends with the deflection angle $\alpha$ near a lens object $L$, i.e. a wormhole.
It is observed by an observer O as an image $I$ with an image angle $\theta$ but not a source angle $\phi$.
$b$ is the impact parameter.
The effective deflection angle $\bar{\alpha}$ is defined as $\bar{\alpha} \equiv \alpha -  2\pi N$,
where $N$ is a non-negative integer which denotes the winding number of the light ray. 
$D_{OL}$, $D_{LS}$ and $D_{OS}$ are the distances from the observer to the lens, from the lens to the source, and from the observer to the source, respectively.
}
\label{Lens_Configuration}
\end{figure}

We assume that the source and the observer are far away from the lens, 
\begin{equation}\label{eq:far_away}
D_{OL},\; D_{LS} \gg b, 
\end{equation}
where $D_{OL}$ and $D_{LS}$ are the distances from the observer to the lens and from the lens to the source, respectively.
Under the assumption~(\ref{eq:far_away}), 
we obtain
$\left| \phi \right|$, $ \left| \bar{\alpha} \right| \ll 1$,
\begin{equation}\label{eq:b}
b=D_{OL}\theta
\end{equation}
and 
$a/D_{OL}< \theta \ll 1$.
We also assume a thin-lens approximation that light rays only bend on a lens plane.

We use a small angle lens equation~\cite{Bozza:2008ev} given by
\begin{equation}\label{eq:Lens}
D_{LS}\bar{\alpha}=D_{OS}(\theta-\phi),
\end{equation}
where $D_{OS}=D_{OL}+D_{LS}$ is the distance between the observer and the source.
The lens equation always has a positive solution $\theta_{N}(\phi)$ for every non-negative $N$.
We define the magnification $\mu(\phi)$ of an image with an image angle $\theta$ as
\begin{equation}\label{eq:magnification}
\mu(\phi) \equiv \frac{\theta}{\phi} \frac{d\theta}{d\phi}.
\end{equation}

We have considered the case $b>a$ or $\theta>a/D_{OL}$ above. 
In the case for $b<-a$ or $\theta<-a/D_{OL}$, 
we can obtain a negative solution $\theta=\theta_{-N}(\phi)$ of the small angle lens equation for every non-negative $N$.
From spherical symmetry, we obtain  
\begin{equation}\label{eq:Negative_angle}
\theta_{-N}(\phi)=-\theta_{N}(-\phi)
\end{equation}
and
\begin{equation}\label{eq:Negative_magnification}
\mu_{-N}(\phi)=\mu_{N}(-\phi),
\end{equation}
where $\mu_{N}(\phi)$ and $\mu_{-N}(\phi)$ 
are the magnifications of the images with the image angles $\theta_{N}(\phi)$ and $\theta_{-N}(\phi)$, 
respectively.

\section{Gravitational lensing in the strong deflection limit}
We will calculate image angles and magnifications in the Ellis wormhole spacetime based on Bozza's method~\cite{Bozza_2002}.
Using Eq.~(\ref{eq:b}),
the deflection angle~(\ref{eq:strong_field_limit}) in the strong deflection limit can be expressed by
\begin{eqnarray}\label{eq:alpha_theta}
\alpha(\theta)
&=&-\log \left( \frac{\theta}{\theta_{\infty}}-1 \right) + 3\log 2 -\pi \nonumber\\
 &&+O\left( (\theta -\theta_{\infty})\log (\theta -\theta_{\infty})  \right),
\end{eqnarray}
where 
\begin{equation}\label{eq:def_theta_infty}
\theta_{\infty}\equiv \frac{b_{c}}{D_{OL}}=\frac{a}{D_{OL}}.
\end{equation}

For a positive winding number $N\geq 1$,
we define an angle $\theta_{N}^0$ as
\begin{equation}\label{eq:def_theta_N0}
\alpha(\theta_{N}^0)=2\pi N.
\end{equation}
From Eqs.~(\ref{eq:alpha_theta}) and (\ref{eq:def_theta_N0}), $\theta_{N}^0$ is obtained as
\begin{eqnarray}\label{eq:theta_N0}
\theta_{N}^0
=\left(1+\frac{8}{e^{(1+2N)\pi}} \right)\theta_{\infty}.
\end{eqnarray}
We expand $\alpha(\theta)$ around $\theta=\theta_{N}^0$,
\begin{eqnarray}\label{eq:alpha_expand}
\alpha(\theta)
&=&\alpha(\theta_{N}^0)+\left. \frac{d\alpha}{d\theta} \right|_{\theta=\theta_{N}^0} \left( \theta-\theta_{N}^0 \right) \nonumber\\
&&+O((\theta-\theta_{N}^0)^{2}).
\end{eqnarray}
When $\theta=\theta_{N}(\phi)$, where $\theta_{N}(\phi)$ is the solution of the lens equation~(\ref{eq:Lens}),
from Eqs.~(\ref{eq:deflection_angle_separeted}), (\ref{eq:Lens}), (\ref{eq:alpha_theta}), and (\ref{eq:def_theta_N0})-(\ref{eq:alpha_expand}), 
we obtain the effective deflection angle as
\begin{equation}\label{eq:effective_deflection_angle_solu}
\bar{\alpha}=-\frac{e^{(1+2N)\pi}}{8\theta_{\infty}}\Delta \theta_{N},
\end{equation}
where $\Delta \theta_{N}$ is defined as
\begin{equation}\label{eq:delta_theta}
\Delta \theta_{N}\equiv \theta_{N}(\phi)-\theta_{N}^0.
\end{equation}

From Eqs.~(\ref{eq:Lens}), (\ref{eq:effective_deflection_angle_solu}) and (\ref{eq:delta_theta}),
we obtain the image angle $\theta_{N}(\phi)$ as 
\begin{equation}\label{eq:theta_N}
\theta_{N}(\phi)
\sim \theta_{N}^{0} + 8\frac{D_{OS}}{D_{LS}}\frac{\theta_{\infty}\left( \phi-\theta^{0}_{N} \right)}{e^{(1+2N)\pi}}.
\end{equation}
In the limit $N\rightarrow \infty$, from Eqs.~(\ref{eq:def_theta_infty}), (\ref{eq:theta_N0}), and (\ref{eq:theta_N}), we obtain
\begin{equation}
\lim_{N\rightarrow \infty}\theta_{N}(\phi)=\theta^{0}_{\infty}=\theta_{\infty}=\frac{b_{c}}{D_{OL}}=\frac{a}{D_{OL}}.
\end{equation}
Thus, $\theta_{\infty}$ is the solution of the lens equation with $N \rightarrow \infty$, 
i.e., the image angle of a light ray that winds around the wormhole throat countable infinite times and then reaches the observer.
It is the innermost image among countable infinite images.

In a perfect alignment $\phi=0$, 
an Einstein ring appears for every positive $N$ because of spherical symmetry.
From Eq.~(\ref{eq:theta_N}), the image angle of the Einstein ring is given by
\begin{equation}\label{eq:Einstein_ring_N_positive}
\theta_{N}(0)\sim \left( 1-8\frac{D_{OS}}{D_{LS}}\frac{\theta_{\infty}}{e^{(1+2N)\pi}} \right) \theta_{N}^{0}.
\end{equation}

From Eqs.~(\ref{eq:magnification}), (\ref{eq:theta_N0}), and (\ref{eq:theta_N}), the magnification $\mu_{N}(\phi)$ is given by
\begin{equation}\label{eq:mu_n}
\mu_{N}(\phi)
\sim \frac{8}{\phi}\frac{D_{OS}}{D_{LS}}\frac{\theta_{\infty}^{2}}{e^{(1+2N)\pi}}\left( 1+\frac{8}{e^{(1+2N)\pi}} \right).
\end{equation}
The sums of the magnifications of the images with $N\geq 1$ and $N\geq 2$ are given by
\begin{eqnarray}\label{eq:sum_mu_1}
\sum^{\infty}_{N=1}\mu_{N}(\phi)
\sim \mu_{1}(\phi)
\sim \frac{8}{\phi}\frac{D_{OS}}{D_{LS}}\frac{\theta_{\infty}^{2}}{e^{3\pi}}
\end{eqnarray}
and 
\begin{eqnarray}\label{eq:sum_mu_2}
\sum^{\infty}_{N=2}\mu_{N}(\phi)
\sim \mu_{2}(\phi)
\sim \frac{8}{\phi}\frac{D_{OS}}{D_{LS}}\frac{\theta_{\infty}^{2}}{e^{5\pi}},
\end{eqnarray}
respectively.

If we observe the image with $N=1$ and the inner images with $N\geq 2$ separately, 
from Eqs.~(\ref{eq:theta_N0}), (\ref{eq:theta_N}), (\ref{eq:sum_mu_1}) and (\ref{eq:sum_mu_2}),
we can measure 
\begin{equation}\label{eq:s_obs}
s_{\mathrm{obs}}\equiv \theta_{1}(\phi)-\theta_{2}(\phi)
\sim \theta_{1}(\phi)-\theta_{\infty}
\sim \frac{8}{e^{3\pi}}\theta_{\infty}
\end{equation}
and
\begin{eqnarray}\label{eq:ratio}
r_{\mathrm{obs}}\equiv 
\frac{\mu_{1}(\phi)}{\sum^{\infty}_{N=2}\mu_{N}(\phi)}
\sim e^{2\pi}.
\end{eqnarray}
From the observables $s_{\mathrm{obs}}$ and $r_{\mathrm{obs}}$, we can check whether the lens object is an Ellis wormhole or not.

From Eqs.~(\ref{eq:Negative_angle}), (\ref{eq:Negative_magnification}), (\ref{eq:theta_N}) and (\ref{eq:mu_n}),
we get the image angle $\theta_{-N}(\phi)$ and the magnification $\mu_{-N}(\phi)$ of the other image 
\begin{equation}
\theta_{-N}(\phi)
\sim -\theta_{N}^{0} + 8\frac{D_{OS}}{D_{LS}}\frac{\theta_{\infty}\left( \phi+\theta^{0}_{N} \right)}{e^{(1+2N)\pi}}
\end{equation}
and 
\begin{equation}
\mu_{-N}(\phi)
\sim -\frac{8}{\phi}\frac{D_{OS}}{D_{LS}}\frac{\theta_{\infty}^{2}}{e^{(1+2N)\pi}}\left( 1+\frac{8}{e^{(1+2N)\pi}} \right),
\end{equation}
respectively.
Thus, the total magnification $\mu_{N \mathrm{tot}(\phi)}$ of both images is given by
\begin{eqnarray}\label{eq:total_magnification_N}
\mu_{N \mathrm{tot}}(\phi)
&\equiv& \left| \mu_{N}(\phi) \right| + \left| \mu_{-N}(\phi) \right| \nonumber\\
&\sim& \frac{16}{\left| \phi \right|}\frac{D_{OS}}{D_{LS}}\frac{\theta_{\infty}^{2}}{e^{(1+2N)\pi}}
\end{eqnarray}
for each positive $N$.
The sum of the total magnification of both images of all positive $N$ is obtained as
\begin{equation}\label{eq:sum_of_total_magnification_N}
\sum^{\infty}_{N=1} \mu_{N \mathrm{tot}}(\phi)
\sim \mu_{1 \mathrm{tot}}(\phi) 
\sim \frac{16}{\left| \phi \right|}\frac{D_{OS}}{D_{LS}}\frac{\theta_{\infty}^{2}}{e^{3\pi}}.
\end{equation}

\section{An effect of images with $N\geq 1$ on light curves}
\subsection{Weak-field approximation}
We review  very briefly gravitational lenses by the Ellis wormhole under the weak-field approximation, 
\begin{equation}
a \ll b \qquad  \mathrm{or} \qquad  k \ll  1.
\end{equation}
Under the weak-field approximation, from Eq.~(\ref{eq:weak_deflection}), the winding number vanishes, $N=0$, 
and then the effective deflection angle becomes 
\begin{equation}\label{eq:weak_deflection3}
\bar{\alpha}=\alpha \sim \frac{\pi}{4} k^{2}.
\end{equation}

In a perfect alignment $\phi=0$, an Einstein ring appears with an image angle $\theta=\theta_{0}\equiv \theta_{0}(0)$ from spherical symmetry.
From Eqs.~(\ref{eq:b}), (\ref{eq:Lens}) and (\ref{eq:weak_deflection3}),  
the image angle $\theta_{0}$ of the Einstein ring is obtained as
\begin{equation}\label{eq:Einstein_ring_N0}
\theta_{0}\equiv \left( \frac{\pi D_{LS} a^{2}}{4D_{OS}D_{OL}^{2}} \right)^{\frac{1}{3}}.
\end{equation}

In an almost alignment case $\left| \phi \right| \ll \theta_{0}$,
the image angle $\theta_{0}(\phi)$ and the magnification $\mu_{0}(\phi)$ are
\begin{equation}\label{eq:image_angle_weak}
\theta_{0}(\phi)\sim \theta_{0}+\frac{1}{3}\phi +\frac{\phi^{2}}{9\theta_{0}}
\end{equation}
and 
\begin{equation}\label{eq:magnification_weak}
\mu_{0}(\phi)
\sim \frac{1}{3}\left( \frac{\theta_{0}}{\phi}+1 \right),
\end{equation}
respectively.
From Eqs.~(\ref{eq:Negative_angle}), (\ref{eq:Negative_magnification}), (\ref{eq:image_angle_weak}), and (\ref{eq:magnification_weak}), 
the image angle $\theta_{-0}(\phi)$ and the magnification $\mu_{-0}(\phi)$
of the other image are obtained as
\begin{equation}
\theta_{-0}(\phi)\sim -\theta_{0}+\frac{1}{3}\phi -\frac{\phi^{2}}{9\theta_{0}}
\end{equation}
and 
\begin{equation}
\mu_{-0}(\phi)
\sim \frac{1}{3}\left( -\frac{\theta_{0}}{\phi}+1 \right),
\end{equation}
respectively.
The total magnification $\mu_{0\mathrm{tot}}(\phi)$ is given by
\begin{equation}\label{eq:magnification_weak_tot}
\mu_{0\mathrm{tot}}(\phi)
\equiv \left| \mu_{0}(\phi) \right| +\left| \mu_{-0}(\phi) \right|
\sim \frac{2\theta_{0}}{3\left| \phi \right|}.
\end{equation}

\subsection{An effect of images with $N\geq 1$ on light curves}
We consider an effect of images with $N\geq 1$ on microlensing light curves in the Ellis wormhole spacetime. 
In an almost alignment case $\left| \phi \right| \ll \theta_{0}$,
from Eqs.~(\ref{eq:def_theta_infty}), (\ref{eq:sum_of_total_magnification_N}), (\ref{eq:Einstein_ring_N0}), and  (\ref{eq:magnification_weak_tot}),
the ratio $R$ of $\sum^{\infty}_{N=1}\mu_{N\mathrm{tot}}(\phi)$ divided by $\mu_{0\mathrm{tot}}(\phi)$
is obtained as
\begin{equation}\label{eq:delta}
R
\equiv \frac{\sum^{\infty}_{N=1}\mu_{N\mathrm{tot}}(\phi)}{\mu_{0\mathrm{tot}}(\phi)} 
\sim \frac{2^{\frac{11}{3}}3}{e^{3\pi}\pi^{\frac{1}{3}}} \left( \frac{D_{OS}}{D_{LS}}\theta_{\infty} \right)^\frac{4}{3}.
\end{equation}
On the other hand, the ratio $R$ in the Schwarzschild spacetime is given by~\cite{Tsukamoto:2014dta,Bozza:2001xd}
\begin{equation}\label{eq:delta6}
R
\sim \frac{2^{3} 3^{\frac{15}{4}}(7-4\sqrt{3})}{e^{\pi}} \left( \frac{D_{OS}}{D_{LS}}\theta_{\infty} \right)^\frac{3}{2}.
\end{equation}
We notice that the ratio $R$ is proportional to $(D_{OS}\theta_{\infty}/D_{LS})^\frac{4}{3}$ in the Ellis wormhole spacetime but
to $(D_{OS}\theta_{\infty}/D_{LS})^\frac{3}{2}$ in the Schwarzschild spacetime.

As discussed in Ref.~\cite{Abe_2010},
we consider 
a bulge star at $D_{OS}=8$kpc and an Ellis wormhole with $1 \mathrm{km} \leq  a \leq 1\times 10^{11} \mathrm{km}$ at $D_{OL}=4$kpc 
and a star in the Large Magellanic Cloud at $D_{OS}=50$kpc and an Ellis wormhole at $D_{OL}=25$kpc.
See Tables~\ref{table:I} and~\ref{table:II} for the two cases. 
Since $R$ is tiny, we conclude that we can ignore the effect of the images with $N\geq 1$ on light curves in both cases.
\begin{table}[hbtp]
 \caption{Gravitational lensing of a bulge star at $D_{OS}=8$kpc by an Ellis wormhole at $D_{OL}=4$kpc.}
 \label{table:I}
\begin{center}
\begin{tabular}{c c c c} \hline
$a$(km) &$\theta_{0}$(mas) &$\theta_{\infty}$(mas) &$R$ \\ \hline
$1.0                    \quad   $ & $ 6.1\times 10^{-4}    \quad   $ & $ 1.7\times 10^{-9}    \quad   $ & $ 8.6\times 10^{-26}$ \\ \hline
$1.0\times 10^{1}       \quad   $ & $ 2.8\times 10^{-3}    \quad   $ & $ 1.7\times 10^{-8}    \quad   $ & $ 1.9\times 10^{-24}$ \\ \hline
$1.0\times 10^{2}       \quad   $ & $ 1.3\times 10^{-2}    \quad   $ & $ 1.7\times 10^{-7}    \quad   $ & $ 4.0\times 10^{-23}$ \\ \hline
$1.0\times 10^{3}       \quad   $ & $ 6.1\times 10^{-2}    \quad   $ & $ 1.7\times 10^{-6}    \quad   $ & $ 8.6\times 10^{-22}$ \\ \hline
$1.0\times 10^{4}       \quad   $ & $ 2.8\times 10^{-1}    \quad   $ & $ 1.7\times 10^{-5}    \quad   $ & $ 1.9\times 10^{-20}$ \\ \hline
$1.0\times 10^{5}       \quad   $ & $ 1.3                  \quad   $ & $ 1.7\times 10^{-4}    \quad   $ & $ 4.0\times 10^{-19}$ \\ \hline
$1.0\times 10^{6}       \quad   $ & $ 6.1                  \quad   $ & $ 1.7\times 10^{-3}    \quad   $ & $ 8.6\times 10^{-18}$ \\ \hline
$1.0\times 10^{7}       \quad   $ & $ 2.8\times 10^{1}     \quad   $ & $ 1.7\times 10^{-2}    \quad   $ & $ 1.9\times 10^{-16}$ \\ \hline
$1.0\times 10^{8}       \quad   $ & $ 1.3\times 10^{2}     \quad   $ & $ 1.7\times 10^{-1}    \quad   $ & $ 4.0\times 10^{-15}$ \\ \hline
$1.0\times 10^{9}       \quad   $ & $ 6.1\times 10^{2}     \quad   $ & $ 1.7                  \quad   $ & $ 8.6\times 10^{-14}$ \\ \hline
$1.0\times 10^{10}      \quad   $ & $ 2.8\times 10^{3}     \quad   $ & $ 1.7\times 10^{1}     \quad   $ & $ 1.9\times 10^{-12}$ \\ \hline
$1.0\times 10^{11}      \quad   $ & $ 1.3\times 10^{4}     \quad   $ & $ 1.7\times 10^{2}     \quad   $ & $ 4.0\times 10^{-11}$ \\ \hline
\end{tabular}
\end{center}
\end{table}
\begin{table}[hbtp]
 \caption{Gravitational lensing of a star in the Large Magellanic Cloud at $D_{OS}=50$kpc by an Ellis wormhole at $D_{OL}=25$kpc.}
 \label{table:II}
\begin{center}
\begin{tabular}{ c c c c} \hline
$a$(km) &$\theta_{0}$(mas) &$\theta_{\infty}$(mas) &$R$ \\ \hline
$1.0                     \quad   $ & $ 1.8\times 10^{-4}    \quad   $ & $ 2.7\times 10^{-10}   \quad   $ & $ 7.5\times 10^{-27}$ \\ \hline
$1.0\times 10^{1}        \quad   $ & $ 8.3\times 10^{-4}    \quad   $ & $ 2.7\times 10^{-9}    \quad   $ & $ 1.6\times 10^{-25}$ \\ \hline
$1.0\times 10^{2}        \quad   $ & $ 3.9\times 10^{-3}    \quad   $ & $ 2.7\times 10^{-8}    \quad   $ & $ 3.5\times 10^{-24}$ \\ \hline
$1.0\times 10^{3}        \quad   $ & $ 1.8\times 10^{-2}    \quad   $ & $ 2.7\times 10^{-7}    \quad   $ & $ 7.5\times 10^{-23}$ \\ \hline
$1.0\times 10^{4}        \quad   $ & $ 8.3\times 10^{-2}    \quad   $ & $ 2.7\times 10^{-6}    \quad   $ & $ 1.6\times 10^{-21}$ \\ \hline
$1.0\times 10^{5}        \quad   $ & $ 3.9\times 10^{-1}    \quad   $ & $ 2.7\times 10^{-5}    \quad   $ & $ 3.5\times 10^{-20}$ \\ \hline
$1.0\times 10^{6}        \quad   $ & $ 1.8                  \quad   $ & $ 2.7\times 10^{-4}    \quad   $ & $ 7.5\times 10^{-19}$ \\ \hline
$1.0\times 10^{7}        \quad   $ & $ 8.3                  \quad   $ & $ 2.7\times 10^{-3}    \quad   $ & $ 1.6\times 10^{-17}$ \\ \hline
$1.0\times 10^{8}        \quad   $ & $ 3.9\times 10^{1}     \quad   $ & $ 2.7\times 10^{-2}    \quad   $ & $ 3.5\times 10^{-16}$ \\ \hline
$1.0\times 10^{9}        \quad   $ & $ 1.8\times 10^{2}     \quad   $ & $ 2.7\times 10^{-1}    \quad   $ & $ 7.5\times 10^{-15}$ \\ \hline
$1.0\times 10^{10}       \quad   $ & $ 8.3\times 10^{2}     \quad   $ & $ 2.7                  \quad   $ & $ 1.6\times 10^{-13}$ \\ \hline
$1.0\times 10^{11}       \quad   $ & $ 3.9\times 10^{3}     \quad   $ & $ 2.7\times 10^{1}     \quad   $ & $ 3.5\times 10^{-12}$ \\ \hline
\end{tabular}
\end{center}
\end{table}

\section{Discussion and conclusion}
In this paper, we have calculated a deflection angle in a strong deflection limit by using two methods in an Ellis wormhole spacetime. 
First, we have obtained it as Eq.~(\ref{eq:strong_field_limit}) 
by considering a behavior of the complete elliptic integral of the first kind $K(k)$ in the limit $k\rightarrow 1$.
Second, we have extended a well-known strong deflection limit analysis in Ref.~\cite{Bozza_2002} 
and then we have obtained a deflection angle~(\ref{eq:strong_field_limit2}) that is equal to Eq.~(\ref{eq:strong_field_limit}).
On the other hand, Eqs.~(\ref{eq:strong_field_limit}) and (\ref{eq:strong_field_limit2}) that we have obtained 
are contradicted by Eq.~(55) in \cite{Nandi_Zhang_Zakharov_2006}.
The difference has influence on observables in the strong gravitational field. 

We have obtained an equal order of an error term $O((b-b_{c})\log(b-b_{c}))$
of the deflection angle in the strong deflection limit by using the above two methods but not $O(b-b_{c})$ given in~\cite{Bozza_2002}. 
Which order is correct? 
Iyer and Petters~\cite{Iyer:2006cn} have investigated a strong deflection series expansion in the Schwarzschild spacetime and 
the exact term in the order $O((b-b_{c})\log(b-b_{c}))$ is already known~\cite{Bozza_Private}.
We have noticed that the error term $O((b-b_{c})\log(b-b_{c}))$ of the divergent part of the deflection angle $I_{D}$
was neglected in Eqs.~(19) and (22) in Ref.~\cite{Bozza_2002}. 
Thus, $O((b-b_{c})\log(b-b_{c}))$ is the correct order of the error term.
Since $O((b-b_{c})\log(b-b_{c}))$ is bigger than $O(b-b_{c})$, 
we should read the error term $O(b-b_{c})$ of the deflection angle in the strong deflection limit 
as $O((b-b_{c})\log(b-b_{c}))$ in~\cite{Bozza_2002} and in dozens of papers using the results of Ref.~\cite{Bozza_2002}.
These dozens of papers mainly discuss the images with the winding number $N \geq 1$ lensed by suppermassive black holes at the center of galaxies. 
Our discussion on the error term here would help us to understand them more clearly. 

In this paper, we have improved the strong deflection limit analysis of the deflection angle.
We have pointed out that the strong deflection limit analysis in Ref.~\cite{Bozza_2002} 
does not work in ultrastatic spacetimes like the Ellis wormhole spacetime.
We have introduced a new variable to obtain the deflection angle in the strong deflection limit.
We could define a similar variable for other ultrastatic spacetimes 
to obtain its deflection angle in the strong deflection limit.

We have extended the reliable analysis~\cite{Bozza_2002} only to obtain the deflection angle in the strong deflection limit in the Ellis wormhole spacetime.
Can we really believe that our revised method works well?
We may need to cross-check the deflection angle in the strong deflection limit.
For example, we could use a sophisticated method investigated by Bozza and Scarpetta 
to obtain the deflection angle in the strong deflection limit~\cite{Bozza:2007gt}.
It seems to work well in ultrastatic spacetimes including the Ellis wormhole spacetime.
Further cross-checking of the deflection angle in the strong deflection limit in the Ellis wormhole spacetime leaves us a future work.

Dey and Sen claimed that there are no images with $N\geq 1$ made light rays passing near a light sphere in the Ellis wormhole spacetime
because of absence of the light sphere~\cite{Dey_Sen_2008}.
Their conclusion, however, contradicts this and other papers~\cite{Nandi_Zhang_Zakharov_2006,Bhattacharya:2010zzb,Chetouani_Clement_1984,Nakajima_Asada_2012,
Gibbons_Vyska_2012,Tsukamoto_Harada_Yajima_2012,Ellis_1973,Muller:2008zza,Perlick_2004_Phys_Rev_D,Muller_2004,Ohgami:2015nra,Perlick:2015vta,Tsukamoto:2016zdu}. 
The Ellis wormhole has the light sphere at its throat and the deflection angle of a light ray passing by the light sphere can be arbitrarily large.
It is worth noting that the leading term of the deflection angle under the weak-field approximation in~\cite{Dey_Sen_2008} 
is the same as the one in Eq.~(\ref{eq:weak_deflection}), i.e, $\pi k^{2}/4$.
Abe~\cite{Abe_2010} and Toki \textit{et al.}~\cite{Toki_Kitamura_Asada_Abe_2011}
discussed the microlenses and the astrometric image centroid displacements in the Ellis wormhole spacetime under the weak-field approximation, respectively, 
by using the leading term of the deflection angle derived by Dey and Sen~\cite{Dey_Sen_2008}.
Their calculations of microlenses and the astrometric image centroid displacements in~\cite{Abe_2010,Toki_Kitamura_Asada_Abe_2011}
are valid 
because they disregarded the effects of the higher-order terms on them.
Recently, Lukmanova~\textit{et al.}~\cite{Lukmanova_2016} have considered 
the effects of a higher term $9\pi k^{4}/64$ in Eq.~(\ref{eq:weak_deflection}) on the light curves.
However, the image angle and image magnification are incorrect because of their miscalculation.
Thus, their higher-order correction on the light curves seems to be doubtful.

Gravitational lenses by the Ellis wormhole are different from the ones by massive objects
since the Ellis wormhole has vanishing ADM masses.
In Ref.~\cite{Abe_2010}, 
Abe found that the total magnification can be less than unity in an Ellis wormhole spacetime 
and the microlensing light curves have gutters near the peak under the weak-field approximation.
Abe concluded that images with $N\geq 1$ do not disturb the characteristic light curves with gutters
since he considered that the Ellis wormhole does not have a light sphere.
In this paper, we have estimated the magnifications of the images with $N\geq 1$ correctly 
as an application of our result on the deflection angle in strong deflection limit in the Ellis wormhole spacetime.
We have found that the images with $N\geq 1$ are much dimmer than images under the weak-field approximation.
Thus, we conclude that the characteristic light curves with gutters are not disturbed by the images with $N\geq 1$.

We emphasize the importance of correct calculations of parameters $\bar{a}$ and $\bar{b}$ in the deflection angle in the strong deflection limit~(\ref{eq:Strong}).
Observables $s_{\mathrm{obs}}$~(\ref{eq:s_obs}), $r_{\mathrm{obs}}$~(\ref{eq:ratio}), and $R$~(\ref{eq:delta}) related light spheres 
strongly depend on $\bar{a}$ and $\bar{b}$. 
Since we have only assumed scattered photons with the impact parameter $b > a$ to obtain $\bar{a}$ and $\bar{b}$, 
one can apply our results to any gravitational lens configurations and any lens equations as long as we consider light rays scattered by a light sphere.
Parameters $\bar{a}$ and $\bar{b}$ of the deflection angle in the strong deflection limit are one of the most fundamental values related to light spheres.
Relations between the parameters $\bar{a}$ and $\bar{b}$ and the quasinormal modes~\cite{Stefanov:2010xz,Wei:2013mda} 
and the high-energy absorption cross section~\cite{Wei:2011zw} have been investigated recently.

\section*{Acknowledgements}
The author thanks V.~Bozza for his encouraging and valuable comments.
The author is grateful to R.~A.~Konoplya for bringing his attention to quasinormal modes.
The author thanks T.~Harada and G.~W.~Gibbons for valuable comments.
He also thanks referees for their valuable comments to correct his misreading of a result in Refs.~\cite{Nandi_Zhang_Zakharov_2006,Bhattacharya:2010zzb}.
The author acknowledges support for this work by the Natural Science Foundation of China under Grant No. 11475065
and the Program for New Century Excellent Talents in University under Grant No. NCET-12-0205.
\appendix

\section{Arnowitt-Deser-Misner mass of the Ellis wormhole}
We can define two ADM masses in the Ellis wormhole spacetime since there are two asymptotically flat regions.
Recently, Nandi~\textit{et al.} pointed out that the Ellis wormhole has nonzero ADM masses 
but the total ADM mass is $0$~\cite{Nandi:2016ccg}.
In this appendix, we clearly show that the Ellis wormhole does have zero ADM masses.
See~\cite{Poisson} for the details of the ADM mass.

We introduce $\rho \equiv \pm \sqrt{r^{2}+a^{2}}$ and rewrite Eq.~(\ref{eq:line_element}) into
\begin{equation}\label{eq:line_element_2}
ds^{2}=-dt^{2}+\frac{d\rho^{2}}{1-\frac{a^{2}}{\rho^{2}}}+\rho^{2}(d\theta^{2}+\sin^{2}\theta d\phi^{2}).
\end{equation}
We consider hypersurfaces $\Sigma_{t}$, which are surfaces of constant t with a unit normal $n_{\alpha}=\partial_{\alpha}t$.
The induced metric on $\Sigma_{t}$ is
\begin{equation}
h_{ab}dy^{a}dy^{b}=\frac{d\rho^{2}}{1-\frac{a^{2}}{\rho^{2}}}+\rho^{2}(d\theta^{2}+\sin^{2}\theta d\phi^{2}).
\end{equation}
The induced metric on a two-sphere $S_{t}$ of $\rho=R_s$ 
with a unit normal $\rho_{a}=\partial_{a}\rho/\sqrt{1-a^{2}/\rho^{2}}$ is given by
\begin{equation}
\sigma_{AB}d\theta^{A}d\theta^{B}=R_s^{2}(d\theta^{2}+\sin^{2}\theta d\phi^{2}).
\end{equation}
The extrinsic curvature of $S_{t}$ embedded in $\Sigma_{t}$ is 
$k=\sigma^{AB}k_{AB}=\rho^{a}_{\;\;  \left| a \right.}=2\sqrt{1-a^{2}/R_s^{2}}/R_s$, 
where $\,_{\left| a \right.}$ denotes the covariant differentiation on $\Sigma_{t}$.
The extrinsic curvature of $S_{t}$ embedded in flat space is $k_{0}=2/R_s$. 

The ADM masses in the Ellis wormhole spacetime are given by
\begin{eqnarray}
M
&=&-\frac{1}{8\pi} \lim_{S_{t}\rightarrow \pm \infty} \oint_{S_{t}}(k-k_{0})\sqrt{\sigma}d^{2}\theta \nonumber\\
&=&\lim_{R_s\rightarrow \pm \infty} R_s\left( 1-\sqrt{1-\frac{a^{2}}{R_s^{2}}} \right)
=0.
\end{eqnarray}
Thus, both its ADM masses vanish.

\section{The deflection angle of a light ray scattered by an Ellis wormhole shown by in previous works}
Equation~(\ref{eq:deflection}) was obtained first by Chetouani and Clement~\cite{Chetouani_Clement_1984}
and later by Nakajima and Asada~\cite{Nakajima_Asada_2012} and by Tsukamoto \textit{et al.}~\cite{Tsukamoto_Harada_Yajima_2012}.

Gibbons and Vyska obtained the deflection angle as Eq.~(136) in \cite{Gibbons_Vyska_2012} 
without the complete elliptic integral of the first kind $K(k)$.
In our notation, Eq.~(136) in \cite{Gibbons_Vyska_2012} is expressed as
\begin{equation}\label{eq:Gibbons}
\alpha=-\pi+\pi\sum^{\infty}_{n=0} \left( {}_{2n} C_{n} \right)^{2} \frac{k^{2n}}{2^{4n}}.
\end{equation}
Using the identity $(2n-1)!!n!=(2n)!2^{-n}$, $(2n)!!=n!2^{n}$, and ${}_{2n} C_{n}=(2n)!/(n!)^{2}$
and
\begin{equation}
K(k)=\frac{\pi}{2} \sum^{\infty}_{n=0} \left[ \frac{(2n-1)!!}{(2n)!!} \right]^{2} k^{2n},
\end{equation}
one can show that the deflection angles Eqs.~(\ref{eq:Gibbons}) and~(\ref{eq:deflection}) are completely the same. 

Dey and Sen obtained a deflection angle in the Ellis wormhole spacetime as
\begin{equation}\label{eq:Dey}
\alpha=-\pi  +\pi \sqrt{\frac{2(r_{0}^{2}+a^{2})}{2r^{2}_{0}+a^{2}}}
\end{equation}
and claimed that the Ellis wormhole does not have a light sphere~\cite{Dey_Sen_2008}.
Under the weak-field approximation, the deflection angle becomes~\cite{Dey_Sen_2008}
\begin{equation}\label{eq:Dey_weak}
\alpha=\frac{\pi a^{2}}{4r_{0}^{2}}-\frac{5\pi a^{4}}{32r^{4}_{0}}+ O\left( \left( \frac{a}{r_{0}}\right)^{6} \right).
\end{equation}
In Ref.~\cite{Gibbons_Vyska_2012}, Gibbons and Vyska pointed that the expansions of Eq.~(\ref{eq:Gibbons})
are completely different from Eq.~(\ref{eq:Dey_weak}).
Nakajima and Asada~\cite{Nakajima_Asada_2012} transformed the deflection angle~(\ref{eq:Dey_weak}) into 
\begin{equation}
\alpha=\frac{\pi}{4}k^{2}+\frac{3\pi}{32}k^{4}+O\left( k^{6} \right)
\end{equation}
and they pointed out that the leading term in Eq.~(\ref{eq:Dey_weak}) is only valid but the subleading terms are not.

Nandi \textit{et al.} obtained a deflection angle Eq.~(54) in Ref.~\cite{Nandi_Zhang_Zakharov_2006} in the Ellis wormhole spacetime.
In our notation, the deflection angle is written as
\begin{equation}\label{eq:Nandi}
\alpha=-\pi+\frac{2\sqrt{a^{2}+r_{0}^{2}}}{r_{0}}K\left( \frac{a}{r_{0}} i \right),
\end{equation}
where $i\equiv \sqrt{-1}$.

In Ref.~\cite{Bhattacharya:2010zzb}, Bhattacharya and Potapov obtained the same deflection angle as Eq.~(\ref{eq:Nandi}).
They obtained the expansions of Eq.~(\ref{eq:Nandi}) in the powers of $a/r_{0}$ as Eq.~(\ref{eq:weak_deflection2})
and in the powers of $k$ as Eq.~(\ref{eq:weak_deflection}).
See Eqs.~(4), (5) and (7) in Ref.~\cite{Bhattacharya:2010zzb}.
They recalculated the deflection angle 
and obtained Eq.~(\ref{eq:deflection})
and the expansion in the power of $k$ as Eq.~(\ref{eq:weak_deflection}).
See Eq.~(9) in Ref.~\cite{Bhattacharya:2010zzb}.
They concluded that Eq.~(\ref{eq:Nandi}) is equivalent to Eq.~(\ref{eq:deflection}) 
since a few leading terms of the expansions in the power of $k$ are the same.

Nakajima and Asada~\cite{Nakajima_Asada_2012} commented that Eq.~(\ref{eq:Nandi}) is apparently different from Eq.~(\ref{eq:deflection})
but it is the same as Eq.~(\ref{eq:deflection}). although they did not show that.
Here we clearly show that Eq.~(\ref{eq:Nandi}) is equivalent to Eq.~(\ref{eq:deflection}).
Table~4 in section 13.~8 in Ref.~\cite{Erdelyi} says
\begin{equation}
K\left( \frac{l}{l'}i \right)=l'K(l),
\end{equation}
where $0<l<1$ and $l'\equiv \sqrt{1-l^{2}}$.
If we set $l=a/\sqrt{a^{2}+r_{0}^{2}}$, 
we obtain
$l'=r_{0}/\sqrt{a^{2}+r_{0}^{2}}$
and 
\begin{equation}\label{eq:Ki}
K\left( \frac{a}{r_{0}}i \right)=\frac{r_{0}}{\sqrt{a^{2}+r_{0}^{2}}}K\left(\frac{a}{\sqrt{a^{2}+r_{0}^{2}}}\right).
\end{equation}
Inserting Eq.~(\ref{eq:Ki}) and $k=a/\sqrt{a^{2}+r_{0}^{2}}$ into (\ref{eq:Nandi}), 
we get Eq.~(\ref{eq:deflection}).
Thus, Eq.~(\ref{eq:Nandi}) is equivalent to Eq.~(\ref{eq:deflection}).

\end{document}